# Performance Deterioration of Deep Learning Models after Clinical Deployment: A Case Study with Auto-segmentation for Definitive Prostate Cancer Radiotherapy

Biling Wang[1,3,^], Michael Dohopolski[1,2,^], Ti Bai[1,2], Junjie Wu[1,2], Raquibul Hannan[1,2], Neil Desai[1,2], Aurelie Garant[1,2], Daniel Yang[1,2], Dan Nguyen[1,2], Mu-Han Lin[1,2], Robert Timmerman[1,2], Xinlei Wang[3,4,*], Steve Jiang[1,2,*]

[1]Medical Artificial Intelligence and Automation Laboratory, University of Texas Southwestern Medical Center, Dallas, Texas, USA
[2]Department of Radiation Oncology, University of Texas Southwestern Medical Center, Dallas, Texas, USA
[3]Department of Statistical Science, Southern Methodist University, Dallas, Texas, USA
[4]Department of Mathematics, University of Texas at Arlington, Dallas, Texas, USA
*Co-Correspondence Authors. Email: xinlei.wang@uta.edu, steve.jiang@utsouthwestern.edu
^ Co-first authors.

## Abstract

**Background** Deep learning (DL)-based artificial intelligence (AI) has made significant strides in the medical domain. There's a rising concern that over time, AI models may lose their generalizability, especially with new patient populations or shifting clinical workflows. Thus we evaluated the temporal performance of our DL-based prostate radiotherapy auto-segmentation model, seeking to correlate its efficacy with changes in clinical landscapes.

**Methods** We retrospectively simulated the clinical implementation of our DL model to investigate temporal performance patterns. Our study involved 1328 prostate cancer patients who underwent definitive radiotherapy between January 2006 and August 2022 at the University of Texas Southwestern Medical Center (UTSW). We trained a U-Net-based auto-segmentation model on data obtained between 2006 and 2011 and tested it on data obtained between 2012 and 2022, simulating the model's clinical deployment starting in 2012. We measured the model's performance using the Dice similarity coefficient (DSC), visualized the trends in contour quality using exponentially weighted moving average (EMA) curves. Additionally, we performed Wilcoxon Rank Sum Test to analyze the differences in DSC distributions across distinct periods, and multiple linear regression to investigate the impact of various clinical factors.

**Findings** During the initial deployment of the model from 2012 to 2014, it exhibited peak performance for all three organs, i.e., prostate, rectum, and bladder. However, after 2015, there was a pronounced decline in the EMA DSC for the prostate and rectum, while the bladder contour quality remained relatively stable. Key factors that impacted the prostate contour quality included physician contouring styles, the use of various hydrogel spacer, CT scan slice thickness, MRI-guided contouring, and the use of intravenous (IV) contrast. Rectum contour quality was notably influenced by factors such as slice thickness, physician contouring styles, and the use of various hydrogel spacers. The quality of the bladder contour was primarily affected by the use of IV contrast.

**Interpretation** Our study highlights the inherent challenges of maintaining consistent AI model performance in a rapidly evolving field like clinical medicine. Temporal changes in clinical practices, personnel shifts, and the introduction of new techniques can erode a model's effectiveness over time. Although our prostate radiotherapy auto-segmentation model initially showed promising results, its performance declined with the evolution of clinical practices. Nevertheless, by integrating updated data and recalibrating the model to mirror contemporary clinical practices, we can revitalize and sustain its performance, ensuring its continued relevance in clinical environments.

**Funding** This study is supported by NIH grants R01CA237269, R01CA254377, and R01CA258987.

## Research in context

### Evidence before this study

We searched across PubMed, IEEE Xplore, and Web of Science using keywords such as "artificial intelligence," "deep learning," "machine learning," "model performance deterioration," "performance change," "performance decrease," "performance shift," "calibration drift," and "medicine," without imposing language restrictions in 2021. Numerous studies were identified that discuss or address issues of deep learning model generalizability; however, most were confined to spatial domain. For example, models trained and validated with data from a single institution often underperform when applied elsewhere. The literature in the medical field tends to overlook the potential for performance degradation of models post-deployment, even when they are used within the same institution. Only three papers marginally acknowledged performance deterioration in their machine learning models, but they did not investigate the causes. Two papers highlighted the importance of ongoing monitoring and updating of AI algorithms in healthcare but fell short of examining how and why model performance may alter over time. There was no dedicated study analyzing the reasons for performance changes in deep learning models after their initial deployment.

### Added value of this study

Our study is pioneering in its demonstration that a model's performance can decline significantly over time. We have identified specific variables that contribute to shifts in data distribution, leading to a decrease in performance. Changes in clinical practices, staffing alterations, and the introduction of novel medical techniques can all contribute to diminishing a model's accuracy. By incorporating insights from post-deployment data, we have biennially updated our model and observed consistent enhancements in performance after each update. Recognizing this caveat is critical for the successful clinical deployment of deep learning models.

### Implications of all the available evidence

The implications of our findings extend beyond the particular model and clinical task investigated in this study; they are relevant to various models used in the medical field. Our research underlines the necessity of establishing protocols for constant monitoring and optimization of these models to maintain their effectiveness and value in patient care across multiple medical settings.

## 1. Introduction

Over the last ten years, artificial intelligence (AI), driven by deep learning (DL) techniques, has achieved remarkable progress, particularly in areas such as computer vision (CV) and natural language processing (NLP), leading to transformative developments across numerous applications. This surge has led to significant enthusiasm in the medical realm, and DL-related medical publications have been growing exponentially since 2015.[1] However, despite the promising prospects of DL in the medical field, its practical deployment remains constrained.[2]

This lack of clinical translation is multifactorial. First, the interpretability of many DL models remains a challenge;[3, 4] therefore, clinicians are rightly skeptical when assessing whether a model can be appropriately applied to patient care.[5-9] This skepticism is not unfounded, as issues of generalizability persist in many DL models.[10-20] For instance, a model trained and validated on data from one institution may fail when implemented at another.[21] Another critical issue, often overlooked in medical literature, is the potential for a model's performance to degrade after initial deployment.[22-26] The decline in model performance post-clinical deployment can often be attributed to data drift, such as variations in imaging acquisition protocols over time within the institution, and evolving practice patterns as new faculty join.[27]

In one of the first clinically-oriented studies evaluating a model's performance, Davis et al.[22] observed a temporal decline in their model's ability to predict acute kidney injury. While they attributed this decline to calibration drift, they did not explore the underlying factors in detail. Similarly, Nestor et al.[25] also noted temporal performance changes when predicting mortality and prolonged length of stay. Clearly, there is a pressing need for further research to explore how and why a DL model's performance may deteriorate over time. In this study, we have observed a temporal decrease in the accuracy of our automated prostate segmentation model. Furthermore, we investigated the potential impact of evolving clinical workflows on this observed decline in model performance. We found that by refreshing the model with recent data, we were able to enhance its accuracy.

## 2. Methods and Materials

We retrospectively simulated the clinical implementation of our DL model to investigate temporal performance patterns. Our study involved 1328 prostate cancer patients who underwent definitive external beam radiotherapy (EBRT) between January 2006 and August 2022 at the University of Texas Southwestern Medical Center (UTSW). We trained a U-Net-based auto-segmentation model on data obtained between 2006 and 2011 and tested it on data obtained between 2012 and 2022, simulating the model's clinical deployment starting in 2012. We measured the model's performance using the Dice similarity coefficient (DSC), visualized the trends in contour quality (DSC) using exponentially weighted moving average (EMA) curves. Additionally, we performed Wilcoxon Rank Sum Test to analyze the differences in DSC distributions across distinct periods, and multiple linear regression to investigate the impact of various clinical factors.

### Dataset

In this single-institutional study approved by our institutional review board, we identified 1480 patients at UTSW diagnosed with prostate cancer and treated with definitive EBRT from January 2006 to August 2022. EBRT treatment regimens included conventional, moderately hypofractionated, or ultra-fractionated radiotherapy, also known as stereotactic body radiotherapy (SBRT). All patients had delineated contours on radiotherapy planning computed tomography (CT) for the prostate, rectum, or bladder. We excluded prostate contours that incorporated the seminal vesicles. To be included, patients were required to have at least a prostate, bladder, or rectum contour. Moreover, patients were excluded if significant artifact was observed. Our final cohort comprised 1,328 patients (Figure 1).

Within the final cohort, 982 had well-defined prostate contours, 1269 had available rectum contours, and 1277 had available bladder contours. One hundred and sixty three (163) patients were treated between 2006 and 2011, 203 patients between 2012 and 2014, 602 between 2015 and 2019, and 360 patients were treated between 2020 and 2022.

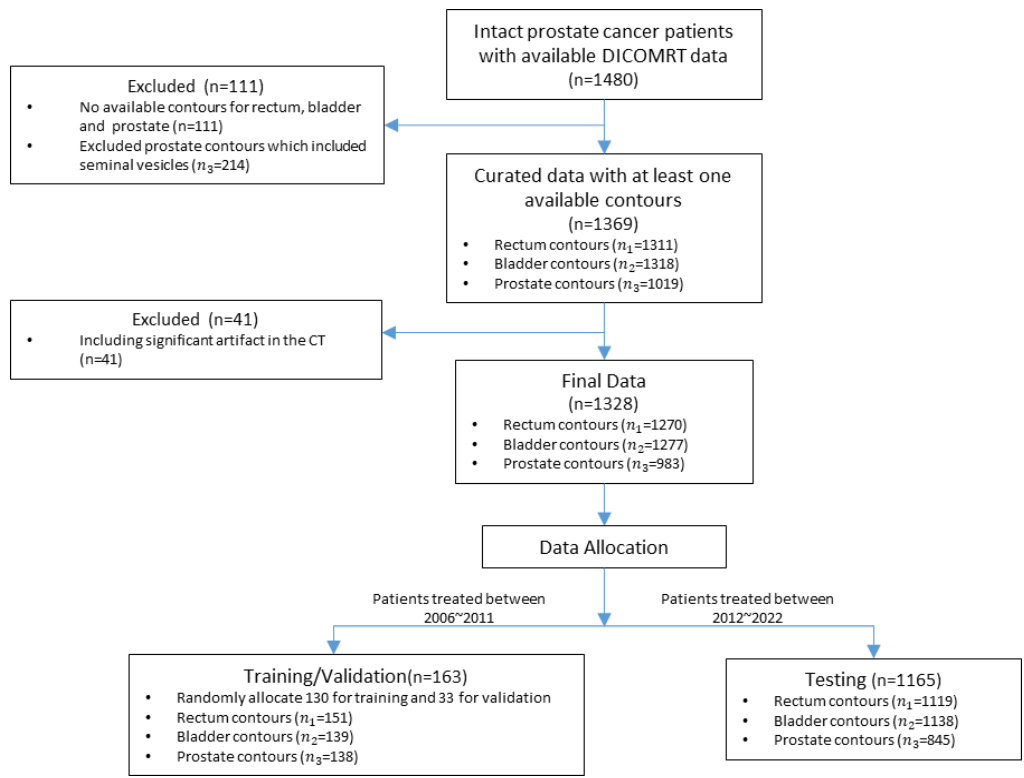

Figure 1. Data selection flow chart

We extracted variables including treating physicians, slice thicknesses for the CT scans, types of hydrogel spacers, either non-contrast (Type I spacer) or contrast-enhancing (Type II spacer), use of intravenous (IV) contrast at time of CT sim (measured by evaluating IV contrast was present in the bladder), and Magnetic Resonance Imaging guided (MRI-guided) contouring techniques. The detailed information of these variables can be seen in section 3.2.

## Model Training

We utilized a 3D U-Net-based auto-segmentation model to contour the prostate, rectum, and bladder on CT images intended for radiotherapy planning. Our implementation was based on the open-source MONAI U-Net.[28] The model was trained using the Adam optimizer with default hyperparameters ($\beta_1 = 0.9$ and $\beta_2 = 0.999$) over $1 \times 10^5$ iterations, leveraging the dice loss function. We initiated the learning rate at $1 \times 10^{-4}$, reducing it to $1 \times 10^{-5}$ at the $4 \times 10^4$th iteration and $1 \times 10^{-6}$ at the $8 \times 10^4$th iteration, respectively. We set the batch size to one. The model was trained and validated using data from 163 patients treated before 2012.

## Longitudinal performance evaluation

DSC was calculated to evaluate the model's performance (i.e. the quality of the contours it generated). The ground truth contours were the clinical contours used in the patients' delivered treatments.

To comprehensively evaluate the model's performance and discern any influencing factors, we employed both visualization techniques and statistical analyses. We charted the model's performance trends from 2012 to 2022 using an EMA. This was applied to the DSC trend curve (termed EMA DSC) with a window size of 180 days and a minimum of 50 observations within the window to have a value. We performed Wilcoxon Rank Sum Test to analyze the differences in DSC distributions across distinct periods.

Our analysis delved into the potential effects of evolving clinical practices on the model's temporal performance. We meticulously examined the influence of factors such as CT imaging slice thickness, types of hydrogel spacers, IV contrast, MRI-guided contouring techniques, and physician styles. We utilized multiple linear regression to identify the key contributors to performance degradation. Since DSC values fell between 0 and 1 and that the DSC distributions of the predicted contours for the prostate, rectum and bladder were left skewed, we applied a logit transformation to the DSC values. Subsequently, these logit-transformed DSC values were fitted into a linear model that incorporated the variables mentioned above and their second-order interactions. We preset a significance level at 0.05 to determine the statistical relevance of each factor. All the analyses were performed using SAS 9.4.

# 3. Results

## 3.1 Model Performance Deterioration

Figure 2A, B and C illustrate the temporal trends in contour quality (DSC) for the prostate, rectum, and bladder. We observed distinct performance patterns for each organ. While the quality of prostate and rectum contours declined significantly over time, bladder contours remained relatively stable.

The model's performance trajectory can be divided into three periods:

- Year 2012-2014: Following its initial deployment, the model exhibited peak performance for all three organs.
- Year 2015-2019: A marked decline in contour quality for the prostate and rectum.
- Year 2020-2022: Modest improvements noted in the quality of the prostate and rectum contours; however, not reaching the quality at initial deployment of the model (2012-2014).

Between 2012-2014 and 2015-2019, the median prostate contour DSC reduced from 0.85 to 0.78 ($p<0.001$), rectum contour DSC reduced from 0.83 to 0.80 ($p<0.001$). The bladder contour exhibited a slight decrease in median DSC from 0.96 to 0.95 ($p=0.003$).

Inspection of the EMA DSC curves corroborated these findings. The prostate's EMA DSC exhibited a pronounced decline—0.84 to 0.74 between 2015 and 2019. Concurrently, the rectum's EMA DSC decreased from 0.82 to 0.76.

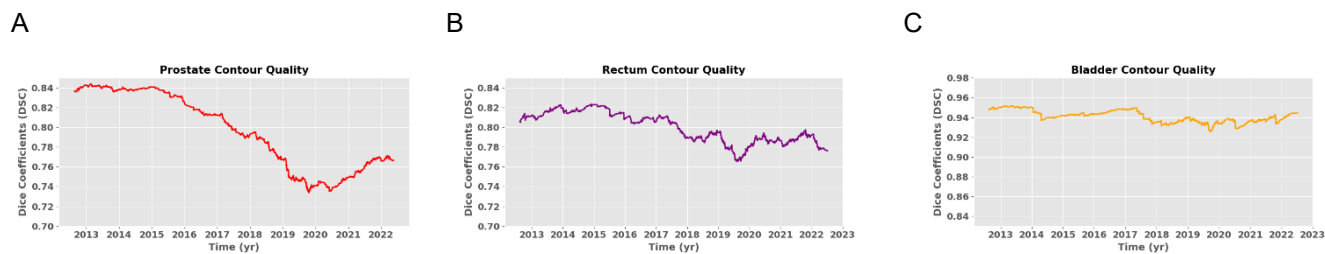

Figure 2. Trends in Auto-Generated Contour Quality

(A), (B) and (C) present the Exponential Weighted Moving Average (EMA) of Dice Similarity Coefficient (EMA DSC) over time post-simulated model deployment for: A) Prostate EMA DSC, B) Rectum EMA DSC, and C) Bladder EMA DSC. EMA DSC for the auto-generated prostate and rectum contours declined, but those for the bladder contours remained stable.

## 3.2 Data Drift

From 2006 to 2011, the majority of patients underwent conventionally fractionated radiotherapy, with a small subset (17.8%) participating in an SBRT clinical trial. From 2012 to 2014, conventionally fractionated radiotherapy remained prevalent, but a significant transition to SBRT was observed post-2015. SBRT, characterized by its precision facilitated by specialized equipment and refined imaging, predominantly utilizes CT scans with a 2 mm slice thickness, compared to the 3 mm thickness common in conventional treatments. The incorporation of hydrogel spacers and IV contrast became more prevalent in SBRT protocols with hydrogel spacers also being used in some patients undergoing moderately hypofractionated treatments. Institutionally, if feasible, MRI-guided prostate contouring was the preferred method for most patients, especially after 2015. As precision radiotherapy gained prominence, these techniques have become foundational in most radiotherapy planning protocols for prostate cancer patients.

Figure 3 presents a comprehensive overview of the data distribution across different time periods, highlighting significant changes over the years. Notably, CT scan slice thickness distribution shifted from a mix of 1.5 mm (n=31, 19.0%), 2 mm (n=32, 19.6%), and 3 mm (n=100, 61.3%) during 2006-2011, to predominantly 2 mm (n=236, 65.6%) and 3 mm (n=124, 34.4%) from 2020-2022. Hydrogel spacer and MRI-guided contouring utilizations were absent in the initial period but were adopted by 84.4% (n=304) and 92.5% (n=360) of patients respectively, during 2020-2022. IV contrast usage rose from 6.7% (n=11) to 41.9% (n=151). A shift in clinical personnel was observed, with Physician A treating 84.7% (n=138) of patients initially, but in the later period (2020-2022), Physicians B (n=128/35.6%), C (n=120/27.8%), and D (n=100/33.3%), collectively treated 96.7% of cases. All the shifts reflected the evolution in treatment approaches and team composition over the years.

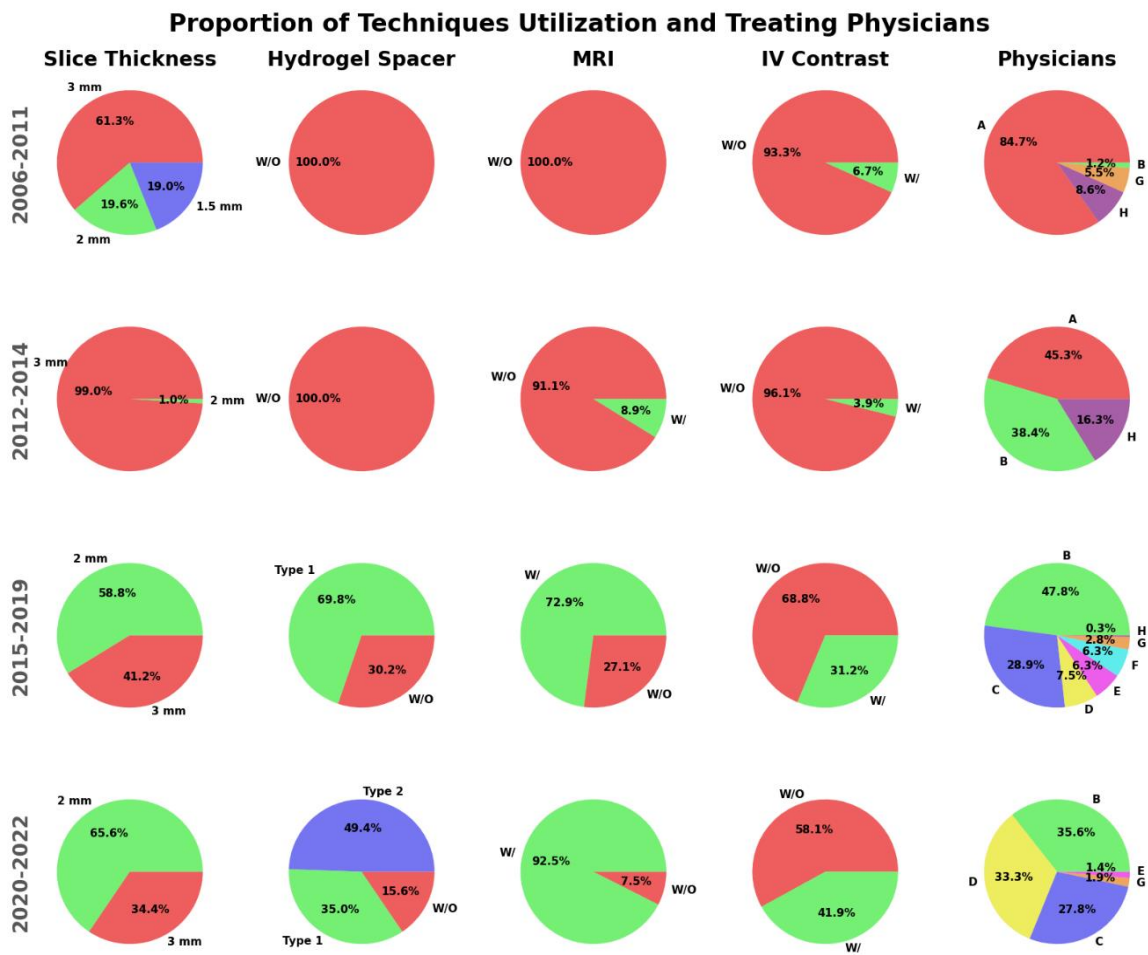

Figure 3. Trends in Data Distribution

Distribution of treatment techniques and physician involvement for the cohorts during year 2006-2011, 2012-2014, 2015-2019, and 2020-2022.

## 3.3 Factors Contributing to the Performance Deterioration

### Impact factors to prostate contour performance deterioration

Figure 4A depicts a decline in the model's prostate contour quality, correlating with the increased incorporation of diverse clinical parameters, including various hydrogel spacer types, slice thickness variations, MRI-guided contouring, and IV contrast usage starting from 2015.

Figure 4B and Table 1 collectively indicate superior prostate contour quality during the 2012-2014 when techniques similar to those used in 2006-2011. Specifically, absence of spacers yielded the highest overall median DSC of 0.84, surpassing type II (0.78) and type I (0.76) spacer groups. A 3 mm slice thickness group exhibited a median DSC of 0.83, outperforming the 2 mm group (0.77). In the context of MRI-guided contouring, the group without this technique registered a superior median DSC of 0.84, compared to the MRI-guided contouring group, which recorded a median DSC of 0.77. Contrastingly, the performance of the IV contrast groups diverged notably during 2020-2022, with the IV contrast group achieving a median DSC of 0.78, surpassing the non-contrast group's 0.76. Additionally, the trend of narrowing differences in subgroup performances suggests potential interaction effects among these variables.

Figure 4C delineates physician-stratified EMA DSC curves, revealing consistent declining trends across subgroups. Physician A's contours were notably superior, establishing a contour quality hierarchy across the evaluated periods: during 2012-2014: A > B; during 2015-2019: B > C and E; during 2020-2022: B > E > C > D.

The multiple linear regression analysis revealed significant associations between prostate contour quality and factors including physician contouring styles, hydrogel spacer types, CT scan slice thickness, MRI-guided contouring, and IV contrast use ($p<0.0001$, $p<0.0001$, $p=0.0085$, $p=0.0012$, $p<0.0001$, respectively, for the main effects only model). We also identified significant interactions between slice thickness and hydrogel spacer types ($p=0.0026$), as well as interactions between physicians' styles and MRI-guided contouring ($p=0.0115$), which impacted prostate contour quality.

### Table 1 Results for Auto-generated Contour Quality

| | Year | Hydrogel Spacer | | | Slice Thickness | | MRI | | IV Contrast | |
|---|---|---|---|---|---|---|---|---|---|---|
| | | W/O | Type I | Type II | 3 mm | 2 mm | W/O | W/ | W/O | W |
| Prostate Median DSC (Q1, Q3)[a] | 2012-2014 | 0.85 (.82, .87) | - | - | 0.85 (.82, .87) | 0.79 (.77, .82) | 0.85 (.82, .87) | 0.84 (.81, .87) | 0.85 (.82, .87) | 0.85 (.77, .87) |
| | 2015-2019 | 0.81 (.77, .85) | 0.76 (.71, .81) | - | 0.81 (.74, .85) | 0.77 (.71, .81) | 0.81 (.76, .85) | 0.77 (.71, .81) | 0.78 (.70, .82) | 0.78 (.75, .82) |
| | 2020-2022 | 0.76 (.72, .82) | 0.75 (.71, .80) | 0.78 (.74, .82) | 0.77 (.71, .82) | 0.77 (.73, .81) | 0.77 (.74, .83) | 0.77 (.73, .81) | 0.76 (.72, .80) | 0.78 (.74, .82) |
| | Overall | 0.84 (.79, .86) | 0.76 (.71, .81) | 0.78 (.74, .82) | 0.83 (.77, .86) | 0.77 (.72, .81) | 0.84 (.80, .86) | 0.77 (.72, .81) | 0.80 (.73, .85) | 0.78 (.75, .82) |
| Rectum Median DSC (Q1, Q3) | 2012-2014 | 0.83 (.79, .87) | - | - | 0.83 (.79, .87) | 0.72 (.69, .75) | 0.83 (.78, .87) | 0.83 (.81, .87) | 0.83 (.79, .87) | 0.82 (.79, .84) |
| | 2015-2019 | 0.82 (.78, .85) | 0.79 (.74, .83) | - | 0.81 (.77, .85) | 0.79 (.73, .83) | 0.82 (.77, .85) | 0.79 (.74, .83) | 0.80 (.75, .84) | 0.80 (.74, .84) |
| | 2020-2022 | 0.79 (.73, .84) | 0.80 (.74, .84) | 0.80 (.75, .84) | 0.81 (.76, .85) | 0.79 (.74, .84) | 0.82 (.73, .87) | 0.80 (.74, .84) | 0.81 (.74, .84) | 0.81 (.75, .84) |
| | Overall | 0.83 (.78, .86) | 0.79 (.74, .83) | 0.80 (.75, .84) | 0.83 (.77, .86) | 0.79 (.74, .83) | 0.83 (.77, .86) | 0.80 (.74, .84) | 0.81 (.75, .85) | 0.81 (.75, .84) |
| Bladder Median DSC (Q1, Q3) | 2012-2014 | 0.96 (.94, .96) | - | - | 0.96 (.94, .96) | 0.93 (.92, 0.95) | 0.96 (.94, .96) | 0.96 (.95, .97) | 0.96 (.94, .96) | 0.93 (.91, .95) |
| | 2015-2019 | 0.95 (.94, .96) | 0.95 (.93, .96) | - | 0.95 (.93, .96) | 0.95 (.93, .96) | 0.95 (.94, .96) | 0.95 (.93, .96) | 0.95 (.94, .96) | 0.95 (.93, .96) |
| | 2020-2022 | 0.95 (.93, .96) | 0.95 (.94, .96) | 0.95 (.94, .96) | 0.95 (.94, .96) | 0.95 (.94, .96) | 0.94 (.92, .96) | 0.95 (.94, .96) | 0.95 (.94, .96) | 0.95 (.94, .96) |
| | Overall | 0.95 (.94, .96) | 0.95 (.93, .96) | 0.95 (.94, .96) | 0.95 (.94, .96) | 0.95 (.94, .96) | 0.95 (.94, .96) | 0.95 (.94, .96) | 0.95 (.94, .96) | 0.95 (.93, .96) |

a, Q1 the first quantile or 25th percentile of the DSC values; Q3 the third quantile or 75th percentile of the DSC values.

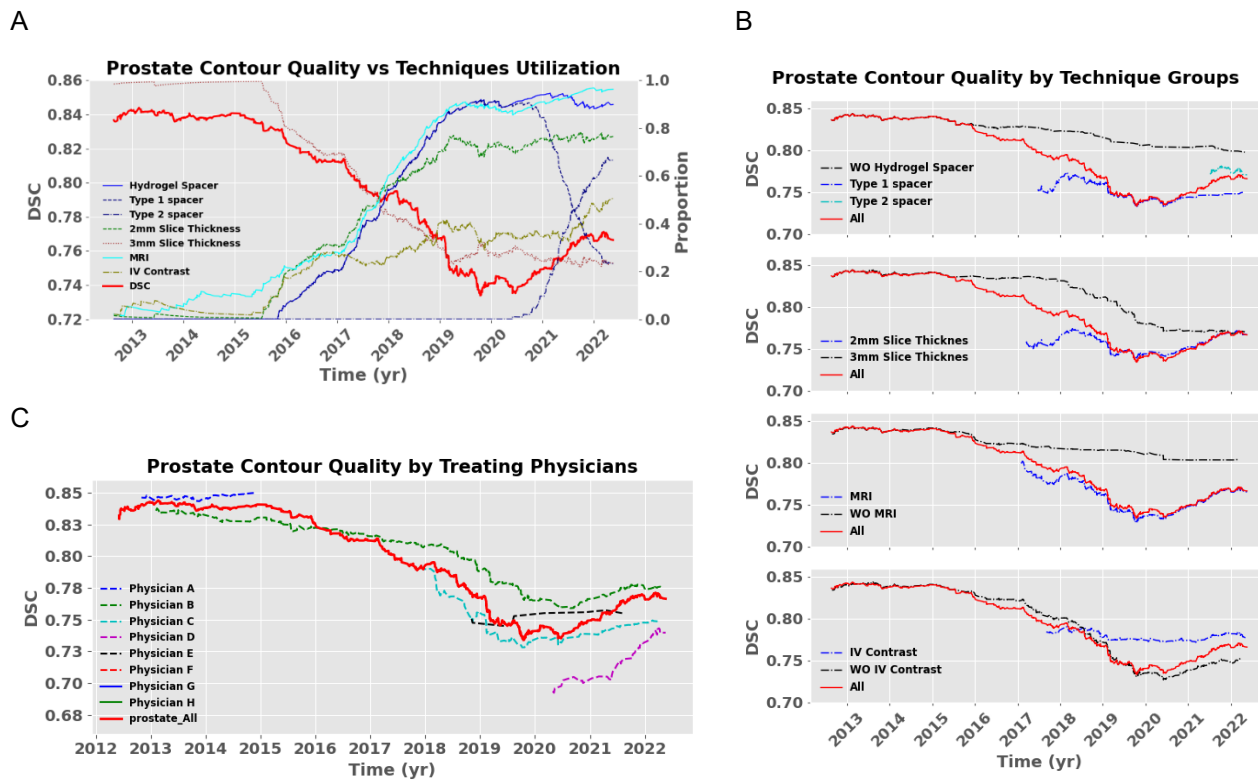

Figure 4. Auto-generated Prostate Contour Quality Vs. Techniques Utilization and Subgroups
(A) EMA DSC for prostate contour quality and proportion of patients with different techniques utilization versus time after the simulated model deployment. (B) Prostate contour quality for subgroups stratified by techniques used versus time. (C) Prostate contour quality for subgroups stratified by treating physicians versus time. The minimum number of observations for the subgroup EMA DSC curves is set to be 30 in figure (C) due to small sample size of patients treated by some physicians. Physicians with available prostate contours fewer than 30 will not have EMA DSC curves

## Impact factors to rectum contour performance deterioration

Figure 5 illustrates trends for rectum contour quality and its subgroups. The rectum contour quality had similar but less pronounced discrepancies for subgroup EMA DSC curves stratified by hydrogel spacer, MRI-guided contouring, slice thickness or IV contrast compared with those of the prostate. Additionally, the rectum contour quality for subgroups maintaining consistent techniques also decreased, indicating no single variable could fully explain the reduction in contour quality.

Figure 5C suggests the model demonstrated superior accuracy for physician B compared to other physicians for generating rectum contours. The contour quality hierarchy was: during 2012-2014: B > H > A; during 2015-2019: H > E > B > C and F; during 2020-2022: E > B > C > D.

The multiple linear regression analysis revealed that rectum contour quality was significantly influenced by factors including slice thickness, physicians' styles, and the use of different types of hydrogel spacers, with p-values of <0.0001, <0.0001, and 0.0470, respectively.

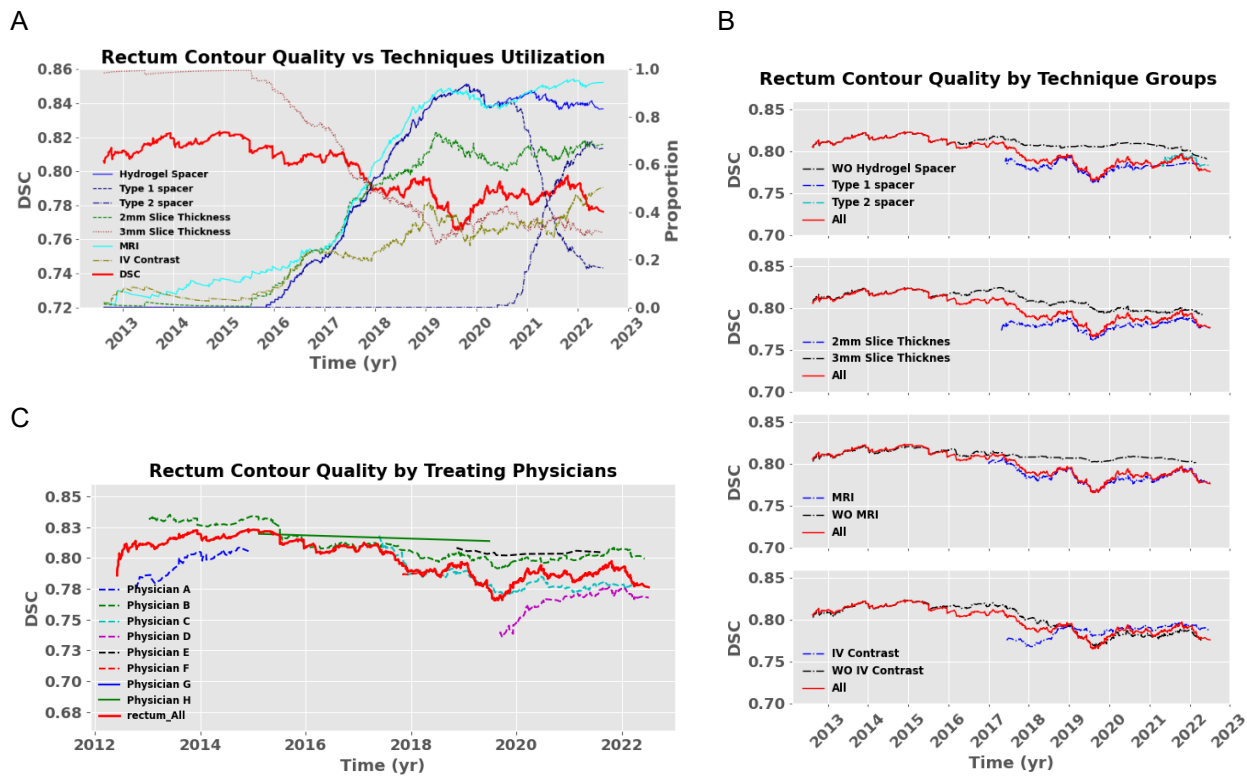

Figure 5. Auto-generated Rectum Contour Quality Vs. Techniques Utilization and Subgroups

(A) EMA DSC for rectum contour quality and proportion of patients with different techniques utilization versus time after the simulated model deployment. (B) Rectum contour quality for subgroups stratified by techniques used versus time. (C) Rectum contour quality for subgroups stratified by treating physicians versus time. The minimum number of observations for the subgroup EMA DSC curves is set to be 30 in figure (C) due to small sample size of patients treated by some physicians. Physicians with available rectum contours fewer than 30 will not have EMA DSC curves.

## Impact factors to bladder contour performance deterioration

Figure 6 underscores the bladder contour quality's stability, demonstrating its resilience to changes in clinical techniques and personnel. The EMA DSC curves of different subgroups are nearly indistinguishable, each maintaining a median DSC of 0.95, indicating consistent performance. However, a deviation is noted between 2012 and 2014, where the IV contrast group exhibited a lower DSC compared to the non-contrast group. Meanwhile, the model showcased consistent efficacy across all physicians (Figure 6C) with minimal variations.

The outcomes of the multiple linear regression analysis indicated the bladder contour quality was primarily influenced by the IV contrast, with a p-value of 0.055.

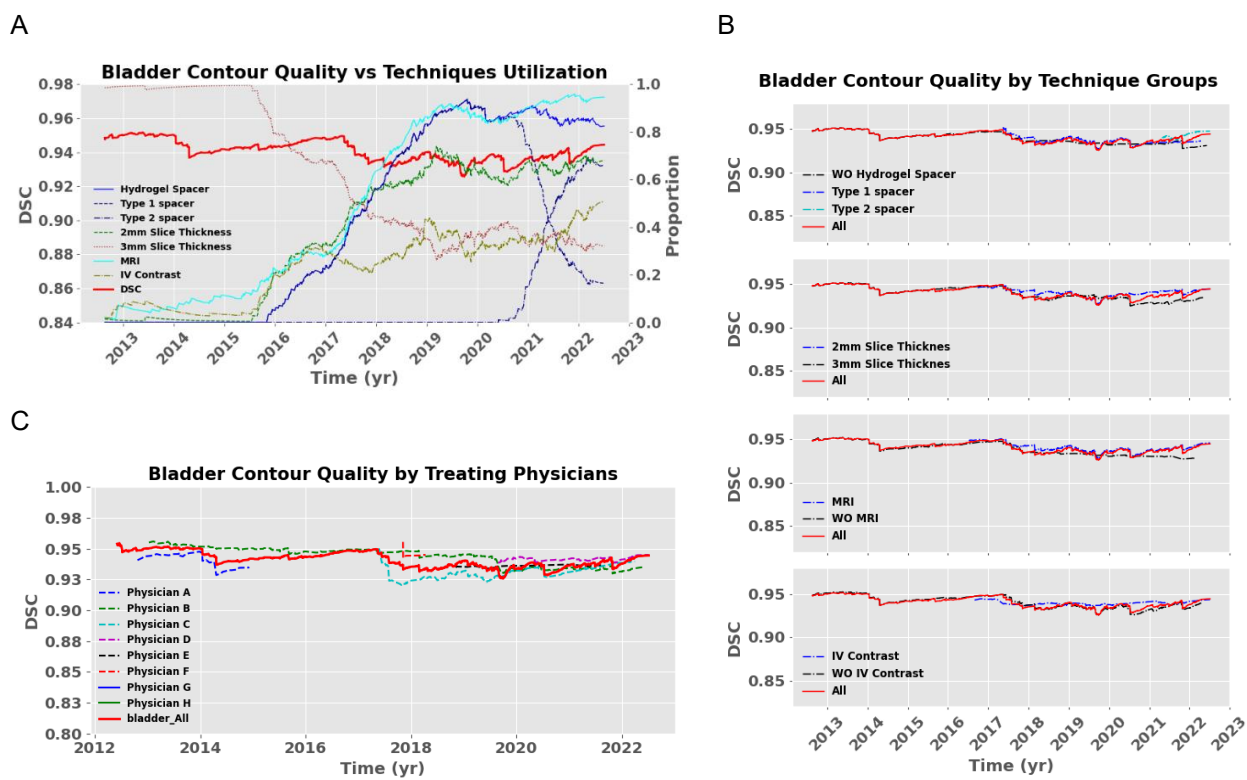

Figure 6. Auto-generated Bladder Contour Quality Vs. Techniques Utilization and Subgroups
(A) EMA DSC for bladder contour quality and proportion of patients with different techniques utilization versus time after the simulated model deployment. (B) Bladder contour quality for subgroups stratified by techniques used versus time. (C) Bladder contour quality for subgroups stratified by treating physicians versus time. The minimum number of observations for the subgroup EMA DSC curves is set to be 30 in figure (C) due to small sample size of patients treated by some physicians. Physicians with available bladder contours fewer than 30 will not have EMA DSC curves.

## 3.4 Model Updates

We found the utilization of the hydrogel spacers with different types, different slice thicknesses, MRI-guided contouring, IV contrast and different physicians' contouring styles are contributors to the DL-based auto-segmentation model's performance deterioration.

To counteract this deterioration, we instigated model updates, incorporating enriched data sets to enhance performance. Specific updates were triggered upon the identification of discernible performance declines and conducted every two years.

- Update Model 1 was trained on data spanning 2006-2016 and tested on data from 2017-2022.
- Update Model 2 utilized training data from 2006-2018 and tested on 2019-2022 data.
- Update Model 3 was trained using data from 2006-2020 and tested on 2021-2022 data.

Figure 7A, B, and C delineate the EMA DSC curves for prostate, rectum, and bladder contours, respectively, illustrating performance enhancements post-update. Each model update manifested in improved contour predictions, although we noticed decline in rectum contour quality in Update Model 2 relative to Model 1 (Figure 7B) in 2019, which was attributed to disparate initial values influencing the EMA curves. Post-2019, the EMA DSC curve for Update Model 1 benefited from the weighted averaging effect of preceding higher values, which was not extended to Update Model 2.

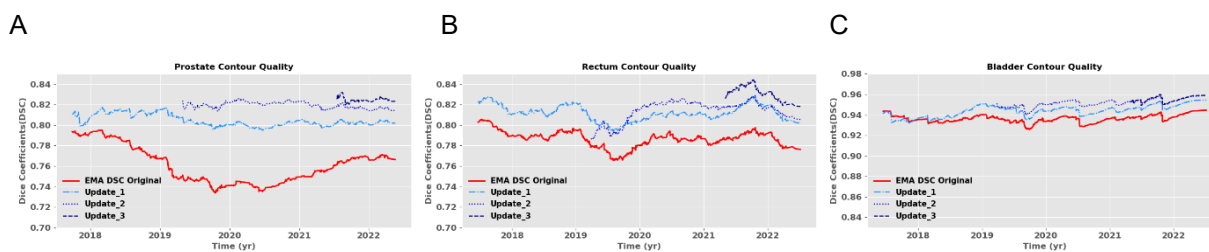

Figure 7. Quality of the model generated contour (EMA DSC) for different models
(A) prostate EMA DSC , (B) rectum EMA DSC, and (C) bladder EMA DSC. The first and second updated model by adding shifted data, the permance for predicting prostate, rectum and bladder contours improved. The third update also improved the model permance for predicting prostate and rectum contour quality.

## 4. Discussion and Conclusions

In this simulated yet practical study evaluating the performance of a DL model deployed in the clinic, we showed that performance decreased over time. We then investigated and identified the potential factors contributing to this deterioration, explicitly evaluating the introduction of new technologies and procedures like hydrogel spacer, CT slice thickness, MRI-guided contouring and the presence of contrast agent in the bladder. Additionally, we considered the changing clinical personnel. To the best of our knowledge, no other studies in the DL auto-contouring medical domain have undertaken a similar exploration into identifying and analyzing factors contributing to the performance deterioration of a deployed DL model over time.

Many DL models published in the literature demonstrate the clinical utility on their specific training and testing datasets.[29-34] However, many of these studies do not evaluate the models' performance years later. Given the evolving of medical practices, shifts in treatment paradigms are inevitable. Such changes can pose challenges for DL models that are not periodically updated for clinical application. Our research revealed our model, trained on data from 2006 to 2011, maintained clinically acceptable performance until the end of 2014. After this period, its ability to contour the prostate and rectum declined significantly. This decline is concerning, especially considering the critical role these structures play in prostate cancer radiotherapy. Radiotherapy targets the prostate while trying to spare adjacent structures, like the rectum and bladder, from receiving toxic radiation doses.[35] An inappropriately contoured target can lead to poor treatment coverage,[36] which increases the patient's risk for recurrence or increases the patient's risk for toxicity if the contoured target inadvertently includes normal tissues, such as the rectum.

It is critical to understand why models might fail in the future. Thus we explored several potential factors that could have impacted our model's performance: the evolution of new treatments and changing personnel. Prostate cancer treatment has evolved over the last several decades. Initially, patients were treated with conventionally fractionated radiotherapy, which included daily radiation treatment, five days a week for more than seven weeks.[37] Over more recent years, patients are more likely to be treated with hypofractionated approaches, which may take only two to four weeks.[38] However, early studies for these hypofractionated approaches did not arise until 2007,[39] with their mainstream clinical adoption occurring much later.

Given this shift, we investigated potential influence of SBRT on our model. Specifically, we assessed the impact of various factors commonly associated with SBRT treatment, such as CT slice thickness, hydrogel spacer usage, MRI-guided contouring, and the use of contrast. Our findings revealed that subgroups utilizing hydrogel spacers (type I and type II), MRI-guided contouring, or 2 mm thick CT slices generally exhibited lower contour quality compared to their counterparts. Specifically, the model's predictions for prostate and rectum contours were inferior in these subgroups compared to those without hydrogel spacers, without MRI-guided contouring, or with 3 mm thick CT slices. Interestingly, the presence of contrast in the bladder enhanced the model's accuracy in predicting prostate contours. However, it's crucial to note that no single technique was solely responsible for the model's performance decline. This is clearly illustrated by the marked decreases observed in the 'without certain technique group' EMA DSC curves (e.g. WO Hydrogel Spacer in Figure 4B), underscoring the multifactorial nature of the performance deterioration.

We quickly observed that the model's ability to contour the rectum and the prostate declined after the type I hydrogel spacer's introduction into the clinic. However, the prostate contour quality improved significantly after type II hydrogel spacer's use in the clinic. In contrast, bladder contour quality remained relatively stable, primarily influenced by the use of IV contrast. Clinically, a hydrogel creates a gap between the prostate and the rectum, reducing rectal radiation exposure and associated side effects.[40] We believe this introduced a data distribution shift by systematically altering traditional anatomy, given that the original model was trained before the FDA's approval of the hydrogel spacer in 2015.[41]

The type I hydrogel spacer, while indistinguishable on CT scans, was visible on MRI images. Consequently, contours for SBRT cases, which predominantly used hydrogel spacers, were delineated with MRI guidance, adhering to our institutional protocol. In contrast, the brighter type II hydrogel spacer was evident on CT scans due to its higher Hounsfield Unit (HU). The pre-trained DL model likely interpreted the 'brighter' type II hydrogel spacer as non-prostate tissue, leading to a noticeable gap between the auto-generated prostate and rectum contours for patients with type II spacer. We speculate that the effect of contrast present within the bladder mirrors that of the type II hydrogel spacer, aiding the model in distinguishing non-prostate tissues. This could explain the enhanced prostate contour quality in patients with contrast present in the bladder. A similar rationale applies to the prediction of rectum contours in the presence of hydrogel spacers and bladder contrast, though the impact of the latter was not statistically significant.

We also found MRI-guided contouring and physicians' contouring styles correlated with the model performance in predicting the prostate and rectum contours. Notably, an interaction effect between physicians' contouring styles and MRI-guided contouring was observed for the prostate contour quality. This suggests that physicians' contours vary depending on whether they utilize MRI imaging or not.

The exact reasoning is unknown, but we postulate that physicians are trained to prioritize target coverage and often contour the entire target based on their expertise and interpretation. While there are established guidelines for contouring the prostate and rectum, the delineation can be ambiguous in certain imaging modalities. MRI-guided contouring, in contrast to CT imaging alone, offers enhanced accuracy and potentially more consistent target and OAR contours.[42] This precision can lead to variations in contouring, especially when combined with individual physician preferences.

Furthermore, it's plausible that the introduction of MRI-guided contouring has influenced physicians to adapt or modify their contouring techniques, given the superior soft tissue contrast and detail provided by MRI. This adaptation might result in contours that deviate from those based solely on CT, leading to discrepancies when compared with DL model predictions trained on CT-based contours.

Additionally, while not widespread, there might be instances where physicians defer OAR contouring to residents or other staff, making only minor adjustments themselves. Such delegation can introduce variability, especially if the training or experience level of the delegates differ from the physicians.

The utilization of a 2 mm thick slice in CT scans, favored for its precision and detailed representation, is predominantly used for SBRT procedures. Notably, the proportion of CT scans employing this 2 mm thickness escalated from 19.6% (training/validation) to 51.2% (testing). However, intriguingly, our observations indicated that the contour quality for both the prostate and rectum was inferior with the 2 mm thick slice compared to the 3 mm thick slice. This counterintuitive finding suggests that while finer slices capture more anatomical details, they might introduce complexities or nuances that challenge the current DL model's ability to generate accurate contours which was trained predominantly on contours based on 3 mm thick CT images.

Other possible explanations for the observed decline in the model's performance need further exploration, especially if the model is to be integrated into clinical practice. This observation underscores the critical need for continuous and systematic monitoring of deployed models. We advocate for the establishment of guidelines that methodically oversee these models' performance, coupled with strategies to investigate contributing variables. Leveraging out-of-distribution quantification metrics, such as uncertainty estimation techniques, might offer a case-specific approach to identify testing instances that diverge from the training dataset.[43, 44]

**Conclusion**

In this novel study, we clearly demonstrated that a model's performance can deteriorate significantly over time. While we could not draw definitive causal conclusions from this retrospective analysis, we pinpointed potential variables that led to shifts in data distribution and subsequent performance decline. Based on these insights, we updated the model biennially using post-deployment data, witnessing consistent performance improvements after each iteration. These findings are not just limited to our specific model but are broadly applicable to other models employed in medicine. Thus, it's crucial to establish guidelines for regular monitoring and refinement of these models, ensuring their continued efficacy and contribution to patient care across various medical domains.

# Contributors

Biling Wang and Michael Dohopolski provided data resources, curated data, searched the literatures, wrote and revised the manuscript. Biling Wang, Michael Dohopolski and Steve Jiang had full access to the data and verified the data. Biling Wang trained the deep learning model and analyzed the data. Ti Bai constructed the deep learning framework and guided the model training. Junjie Wu downloaded all the DICOM data and helped to analyze the DICOM data. Dan Nguyen guided the model training and the analysis. Mu-Han Lin and Michael Dohopolski guided the data curation and contributed essential clinical insights. Raquibul Hannan, Neil Desai, Aurelie Garant, Daniel Yang and Robert Timmerman provided data annotations and clinical expertise. Xinlei Wang guided the overall research study, supervised the statistical evaluation, and contributed editorial assistance to the manuscript. Steve Jiang was pivotal in conceptualizing and directing the research, providing overall guidance on the direction and goals of the project, ensuring the integrity and quality of the research, and contributing editorial assistance to the manuscript. All co-authors have reviewed and consented to the published version of the manuscript.

## Data Sharing Statement

The data that support the findings of this study are available from the corresponding author upon reasonable request. The source code of the deep learning model used in this study are available online and adapted to the investigated data (https://github.com/baiti01/iRT_AutoSegmentation).

## Declaration of interests

We declare no conflict of interests.

## Acknowledgment

This study is supported by NIH grants R01CA237269, R01CA254377, and R01CA258987.
We would like to thank Ms. Sepeadeh Radpour for editing the manuscript.